\documentclass[12pt,preprint]{aastex}

\newcommand{\be}{\begin{equation}}
\newcommand{\ee}{\end{equation}}

\shorttitle{Star formation rate and accretion rate in LINERs }
\shortauthors{Wu \& Cao }

\begin{document}

\title{The relation between star formation rate and accretion rate in LINERs }

\author{$\rm Qingwen\   Wu^{1,\ 2} \ and  \    Xinwu\  Cao^{1}$}

\affil{1. Shanghai Astronomical Observatory, Chinese Academy of
Sciences,
 Shanghai, 200030, China; qwwu@shao.ac.cn, cxw@shao.ac.cn }

\affil{2. Graduate School of Chinese Academy of Sciences, Beijing,
100039, China}

  \clearpage
  \begin{abstract}

    It is argued that there is a linear correlation between
    star formation rate (SFR) and accretion rate for normal bright active galactic
    nuclei (AGNs). However, it is still unclear whether this
    correlation holds for LINERs, of which the accretion rates are
    relatively lower than those of normal bright AGNs. The radiatively
   inefficient accretion flows (RIAFs) are believed to be present in
   these LINERs. In this work, we derive accretion rates for
    a sample of LINERs from their hard X-ray luminosities based on
    spectral calculations for RIAFs. We find that LINERs
    follow the same correlation between star formation rate and
    accretion rate defined by normal bright AGNs, when reasonable
    parameters are adopted for RIAFs. It means that the gases feed the black hole and star formation
     in these low-luminosity LINERs may follow the same way as that in
     normal bright AGNs, which is roughly consistent with recent
    numerical simulations on quasar evolution.

    \end{abstract}

\section{Introduction}
     Black hole (BH) accretion is thought to power AGNs, and the
     UV/optical bump observed in bright quasars is naturally
     interpreted as blackbody emission from standard thin accretion disks \citep*[e.g.,][]{sm89}.
     The AGN activity may be switched off when the gas near the black
     hole is exhausted
     \citep[see][ for a recent review and references therein]{narayan02}. Most nearby galactic nuclei are
      much less active and show very different properties with
     bright quasars,  such as low ionization, lack ``UV bump'' etc \citep*[see][for a
      review]{ho05}. The accretion mode in these nearby galaxies
      may be different from that in powerful AGNs.
      \citet{ny94} proposed that the standard accretion disk should
      transit to a RIAF when the accretion rate $\dot{m}$(=$\dot{M}/\dot{M}_{\rm
      Edd}$) declines below a critical value $\dot{m}_{\rm crit}$
     within a certain transition radius \citep*[e.g.,][]{emn97,yn04,llg04}.
       RIAFs are optically thin, geometrically thick and very hot, which
       are supposed to be present in many low-luminosity
      AGNs \citep*[e.g.,][]{l96,qe99,gnb99,ymfb02,yn04,nt05} and our
      Galactic center Sgr $\rm A^{*}$ \citep*[e.g.,][]{ny95}.

     A very low nuclear-luminosity class of low-ionization nuclear
    emission-line region (LINER) galaxies were identified by \citet*[][]{heckman80},
    in which are found approximately 2/3 low luminosity AGNs
    \citep{hfs97a}. An important fraction of LINERs are clearly weak manifestations of
     quasar-like phenomena, as demonstrated by the presence of broad H$\alpha$ emission, which are almost certainly
     accretion-powered \citep*[][]{ho97b}. These LINERs can be identified with
     quiescent BH remnants from the quasar era. In the present epoch, the supply of the
      gas available for powering central engines is much curtailed.
      The low mass accretion rates inferred for many LINERs may
     suggest that these sources represent the ``missing link'' between
     powerful AGNs and normal galaxies as our own.

     The recent discovery of the correlation between BH mass and stellar velocity dispersion  in
     nearby galaxies \citep*[e.g.,][]{ge00,fm00}  demonstrates a fundamental link
     between the growth of supermassive BH and bulge formation.
     The connection between SFR and accretion rate has been explored
     by some authors \citep*[e.g.,][]{hec04,hao05,sa05,du05}.
     Using the SDSS (Sloan Digital Sky Survey) observations of 123,000 low-redshift
     galaxies, \citet{hec04} found that the global volume-averaged
     SFR/$\dot{M}$ ratio is approximately 1000 in bulge-dominated
     systems, which is in agreement with the ratio of bulge to BH
     mass implied by the $M_{\rm bh}-\sigma$
     relation \citep{mh03}.  \citet{hao05} also yielded similar results for several ten AGNs
     (including infrared-selected QSOs, optically-selected
     QSOs, NLS1s).
     It was found that the ratio SFR/$\dot{M}$ of LINERs
     is apparently different from that of normal bright AGNs, of
     which the mass accretion rate $\dot{M}$ is estimated by
     assuming a constant radiative efficiency for all sources. The
     mass accretion rates $\dot{M}$ derived for LINERs may probably
     be incorrect \citep{sa05}, because RIAFs are believed
     to be in these sources and their radiative efficiencies are
     lower than these for standard thin disks.

      In this paper, we explore whether the LINERs follow the
    same correlation between accretion rate and SFR defined by
    normal bright AGNs. The accretion rates are re-estimated for these LINERs
    based on our spectral calculations for RIAFs.

\section{The sample}
     We adopt the sample given by \citet{sa05} for our present
    investigation, which includes
     33 LINERs with detected hard X-ray nuclear emission (they are
     defined as AGN-LINERs in \citet{sa05}), 54 Seyferts, 15 quasars,
      14 radio galaxies, and 14 narrow-line Seyfert 1s (NLS1s).
     Several LINERs are radio-loud.
     All sources in this
     sample have estimated BH masses, FIR and  bolometric
     luminosities. The BH masses  of these sources are estimated
     through resolved stellar kinematics, reverberation mapping, or
     applying the correlation between optical bulge luminosity and
     central BH mass derived from nearby galaxies (see \citet{sa05}, for more
     details). A significant fraction of galaxies in the sample
     are bulge-dominated and the SFRs of the host galaxies are calculated from $L_{\rm
      FIR}$ by using the calibrated formula, in which the old stars'
      contribution has been properly subtracted
      \citep*[See Eqs. 5 and 6 in][]{sa05}. The calibrated
      formula adopted by \citet{sa05} can also be applied to some early type spiral galaxies
      in the sample, as  this SFR-$L_{\rm
      FIR}$ calibration is suggested to be applicable for all type galaxies \citep{kew02}.
      All data can be found in \citet{sa05}. We note that NGC 3215 is a Wolf-Rayet starburst
       galaxy, which is excluded in our present investigation.
      Some typos of their paper are fixed in this paper \citep*[see,][for erratum]{sa06}.

\section{RIAF spectra}

    In order to calculate the spectrum of a RIAF, we need to obtain the RIAF structure.
     We calculate the global structure of an accretion flow
     surrounding a Schwarzschild BH using a similar approach
     as \citet{nkh97}.
    We adopt a simple $\alpha$-viscosity ($\tau_{r\phi}=\alpha p$)
    and all radiative processes (Synchrotron, Bremsstrahlung and Compton
    scattering) are included consistently in our calculations for RIAF structure.  The advection by
    ions and electrons has been considered in the energy equation, and
    a more realistic state of accreting gas (instead of a
    polytropic index $\gamma_{\rm g}$) is employed in the calculations, which is similar
    to that used by  \citet{m00}. The RIAF structure can be
    calculated when suitable outer boundary conditions are supplied. After
    the RIAF structure is obtained, the spectrum of the flow can be
    calculated in the same way as some previous authors
    \citep*[see,][]{m00,esin96}.

\section{Results}

\subsection{The relation between  $L_{X,\ 2-10\rm\ keV}$  and $\dot{m}$ for RIAF}
     Three-dimensional MHD simulations suggest that the viscosity
     parameter $\alpha$ in the accretion flow is $\sim0.05-0.2$
     \citep{hb02}.
    Recent observations of black hole binaries \citep{ma03} and
    radio-loud AGNs \citep{mcf04}, as well as
    theoretical accretion models \citep*[e.g.,][]{ny95,emn97} suggest that the
    critical accretion rate $\dot{m}_{\rm
    crit}\thicksim0.01$, below which a RIAF is present. Our
    numerical calculations show that $\dot{m}_{\rm
    crit}\thicksim0.01$ requires $\alpha\thicksim0.2$, and we adopt
    $\alpha=0.2$ in all our calculations. The parameter $\beta$ (ratio of gas to
   total pressure in the accretion flow, $\beta=P_{\rm g}/P_{\rm
    tot}$) is not an independent parameter, which can be related with $\alpha$
   as $\beta\approx(6\alpha-3)/(4\alpha-3)$, suggested by MHD simulations \citep*[][]{hgb96,nmq98}. We adopt $\beta\simeq0.8$
   for $\alpha=0.2$.
    Another parameter in our calculations  is the outer radius $R_{\rm
    out}$ of the RIAF. We adopt $R_{\rm out}=200R_{\rm S}$ in all our
    calculations, where the Schwarzschild radius $R_{\rm S}=2GM_{\rm bh}/c^{2}$
    and $M_{\rm bh}$ is the BH mass. How the outer radius $R_{\rm out}$ evolves with $\dot{m}$
   is still unclear, though it may be a function of $\dot{m}$. We
    find that the hard X-ray emission is mainly emitted from the
    inner region of the flow close to the black hole, which is
    insensitive to the value of $R_{\rm out}$ adopted. A typical value $\delta=0.1$ (the fraction
    of the dissipated energy directly heating the
     electrons) is adopted in all our
     calculations.  Now, we can calculate the structure of the RIAF, and then its
     spectrum varying with accretion
     rate $\dot{M}$ when $M_{\rm bh}$ is specified.

   For a fixed BH mass, $M_{\rm bh}=10^{8}M_{\odot}$, our calculation on the
   relation between 2-10 keV X-ray luminosity  $L_{X,\ 2-10\rm
  \ keV}$ and accretion rate (from $\dot{m}=10^{-5}$ up to
   $\dot{m}\sim\dot{m}_{\rm
     crit}\sim10^{-2}$) is plotted in Figure 1. The dependence of 2-10 keV luminosity on $\dot{m}$ can be roughly
     fitted by single power-law (solid line in Fig. 1),
  \be
  \log L_{X,\ 2-10\rm\ keV}= 2.37(\pm0.06)\log
  \dot{m}+46.33(\pm0.22),
  \ee

  For $\dot{m}\lesssim10^{-3.75}$, $L_{X,\ 2-10\rm\ keV}\propto\dot{m}^{2}$ and the 2-10 keV X-ray
  emission is mainly dominated by bremsstrahlung radiation. At higher
  accretion rate, $10^{-3.75}\lesssim\dot{m}\lesssim10^{-3.25}$,
  $L_{X,\ 2-10\rm\ keV}\propto\dot{m}^{3.5}$ and both the Compton and bremsstrahlung radiation is contributes to the
  2-10 keV X-ray emission. The Comptonized component will dominate
  the X-ray emission at higher accretion rate
  $\dot{m}\gtrsim10^{-3.25}$ which have $L_{X,\ 2-10\rm\ keV}\propto\dot{m}^{2.2}$ (see dashed line in Fig. 1).

 Our spectral calculations show that the 2-10 keV X-ray luminosity
  is roughly proportional to the BH mass. So we can
  calculate the accretion rate using the equation (1), if the BH mass and $L_{X,\ 2-10\rm
  \ keV}$ is given.

\subsection{The mass accretion rates of LINERs}
  The central BH masses for all sources in this sample are available,
  so we can re-estimate the accretion rates for these 30 low-luminosity
  LINERs from their
  hard  X-ray luminosity $L_{X,\ 2-10\rm\ keV}$ based on our RIAF spectral
  calculations.
   The remaining three LINERs having X-ray luminosities higher than the
    maximal values expected by the RIAF model calculation.  We estimate
    the accretion rates of these three sources through $\dot{m}=L_{\rm bol}/L_{\rm Edd}$, where $L_{\rm
    bol}$ is derived from the X-ray luminosity by using  $L_{\rm
    bol}=34L_{X,\ 2-10\ \rm keV}$.
   The Eddington ratio of 8 sources in the 14 radio galaxies
   are lower than $10^{-2}$ and these sources may be also in RIAF
   state. Four of these 8 sources (3C 31, 3C 84, 3C 338, 3C 465) have intrinsic
   X-ray luminosity \citep{ev05} and the X-ray emission data for the remaining four sources
  (3C 88, 3C 285, 3C 327, 3C 402) are not available in the literature. We
   convert optical core emission measured by $Hubble\ Space\
   Telescope\ (HST)$ to a corresponding X-ray luminosity
   assuming spectral index $\alpha_{ox}=0.6$ for these four sources \citep{mcf04,fkm04}.
   Following the same way as LINERs, we re-estimate the
   accretion  rates of these radio galaxies. The re-calculated mass
   accretion rates and the \citet{sa05}'s results are listed in Table 1 for comparison.

   In \citet{sa05}, they showed that the distributions of
   SFR/$\dot{M}$ for LINERs and radio galaxies are apparently different from that of
   normal bright AGNs. Their estimates on $\dot{M}$ were carried out by assuming a constant radiative efficiency, which may be
   incorrect for low-luminosity LINERs with RIAFs.
    We plot the distributions of the ratio of SFR to the re-estimated
   mass accretion rate $\dot{M}$ for LINERs and radio galaxies in Figure 2
   (upper and middle panels). We find that the distributions of SFR/$\dot{M}$ for
   LINERs and radio galaxies are
   similar to that of the normal bright AGNs in this sample(Seyfert, Quasar and NLS1, solid lines in Fig.
  2).
   In Figure 3, we plot the relation of SFR and
  $\dot{M}$ for the whole sample using our estimates on mass accretion rates $\dot{M}$.
  The linear regression yields for the normal bright  AGNs
  (Seyfert, Quasar and NLS1):
  \be
   \log SFR=(0.79\pm0.04)\log\dot{M}+(1.66\pm0.06),
  \ee
  where SFR and $\dot{M}$ are both in units of $M_{\odot}\rm
  yr^{-1}$. The linear regression for the whole sample is:
  \be
   \log SFR=(0.73\pm0.04)\log\dot{M}+(1.55\pm0.07).
  \ee
We find that the correlations are similar either for the whole
sample or the subsample of normal bright AGNs (see, Fig. 3). We have
not found any radio galaxies in this sample deviating significantly
from the correlation.

\section{Discussion}

  The bolometric luminosities of AGNs can be estimated
  by integrating their spectral energy distribution or using an empirical relation
  \citep*[e.g.,][]{wu02,el94,ho99}, and then their mass accretion
  rates are derived if the radiative efficiency
  is known. A linear correlation between SFR and $\dot{M}$ was found by
  \citet{sa05} for normal bright AGNs, while the LINERs in their
  sample obviously deviate from this correlation, if the same constant
  radiative efficiency 0.1 is adopted for the whole sample. It was found
  that the accretion rates for LINERs are significantly lower
  than  those of normal bright AGNs for given SFRs \citep*[see Fig. 10
  in][]{sa05}. RIAFs are believed present in these low-luminosity LINERs, and
  their radiative efficiency $\epsilon_{r}$ is usually much lower than
  that of standard disks \citep{ny94}. This implies the accretion
  rates of LINERs are underestimated by \citet{sa05}. In this
  paper, we derive the mass accretion rates of these LINERs from
  their X-ray luminosities based on RIAF spectral calculations,
  which should be more reliable.

  The standard thin disk is believed present
  in Seyferts and QSOs and its radiative efficiency $\epsilon_{r}^{0}\sim0.1$.
  The slim disk may be present in the NLS1s and its radiative
  efficiency should be also lower than that of standard disk due to the photon-trapping effects \citep*[e.g.,][]{ab88,om02,cw04}.
  However, all the Eddington ratios  of NLS1s in our
  sample are less than 6 \citep*[$L_{\rm Bol}/L_{\rm Edd}$, \ see table 2\ in][]{sa05} and the radiative
  efficiencies
   of slim disks with such Eddington ratios may not deviate much from that of the standard
  disk  \citep*[see, dashed line and solid line of Fig. 1 in ][]{om02}.
  So, the derived mass accretion rates should be accurate to a
  factor of 2 for NLS1s in these sample, which will not alter our
  main conclusion.

    The SFRs is calculated from the FIR luminosity for the whole
    sample since the
     FIR luminosity is widely used as a tracer of SFR in galaxies
    \citep*[e.g.,][]{kenn98,kew02,sa05}, though we can not rule out some contribution of
    dust heating by the AGNs and old stars in elliptical galaxies.
    However, there are lack of correlations between the Mid-IR  and FIR
    for well-studied PG quasars, which suggests that they
    are dominated by different heating
    sources\citep*[e.g.,][]{haas99}. If AGN heating
    dominates the cooler FIR-emitting dust, then there should be a
    correlation between quasar OUV and FIR luminosity, while such a correlation has
    not been found\citep*[e.g.,][]{mc99,isaak02,priddey03}. In addition,
   QSOs and Seyferts roughly follow the same universal correlation
   between the FIR and radio emission deduced from ``normal"
   galaxies, which strongly suggests that FIR emission is still
   powered  by star formation rather by AGNs
   \citep*[e.g.,][]{nar88,lw05}.
   Although the FIR emission contributed by old stars in some
   elliptical galaxies  could be important (e.g., very few sources in
   \citet{bell03} sample have old star contribution, about 4 of 249,
   higher 80 percent, see Fig. 6 in that paper), the average
   contribution of the old stars is around 30 percent. In our present
   SFR calculations, we use the calibrated formula, in which the old
   star contribution has been subtracted \citep{bell03}. So, in
   statistical sense, the SFRs estimated in our paper are reliable,
   though we cannot rule out the SFRs of a very few sources in this
   sample have been overestimated 3-4 times. Even in this case, we
   believe this has not affected our main statistical results.

     We plot our re-estimated accretion rate $\dot{M}$ in Figure 3, and
     find that all LINERs follow the same correlation defined by the
     normal bright AGNs (e.g., Seyfert, Quasar and NLS1).
     The linear correlation between SFR and $\dot{M}$ indicates that black
     hole accretion evolves in the same way as star formation, which are
     both regulated by the interstellar gas in the host galaxies. Our
     results show that the gases feed the black hole and star formation in
     low luminosity LINERs follow the similar way as luminous
     normal AGNs. Recent numerical simulations on the quasar activity
     triggered by the galaxy merger indeed show that the accretion rate
     $\dot{M}$ and SFR decreases simultaneously with time over several
     orders of magnitude after the quasar shines at its Eddington limit
     \citep{sdh04,msh05}. In their simulations, the gas in the galaxies
     is driven away by
     the bright quasar radiation, and then both the star formation
      and BH accretion are quenched simultaneously.
     The SFR is nearly linearly varying with $\dot{M}$
    in nearly three orders (e.g., $\dot{M}\thicksim5\times10^{-4}-5\times10^{-1}\ \rm M_{\odot}yr^{-1}$,
   SFR$\thicksim5\times10^{-3}-5 \ \rm M_{\odot}yr^{-1}$ corresponding to
   $\sim1.7-2.0\ \rm Gyr$) for the model with galaxies of virial
     velocity $\rm V_{\rm vir}=160km/s$ \citep*[middle red lines in Fig.
   2 of][]{msh05}. Our results are roughly consistent with their
simulations, though the accretion mode transition has not been
considered in their simulations.

  There are 8 radio galaxies (Eddington ratio less than $10^{-2}$)
   in 14 radio galaxies, of which the accretion
  rates are derived from their X-ray emission. It is still debating
  whether the X-ray emission are dominated by the jet emission for some radio galaxies
  \citep{fkm04,yw05}. If the jet emission in X-ray bands is important, the accretion rates derived in
  this paper are only the upper limits, and those 8 triangles in Figure 3 should
  be shifted towards left direction. As discussed in Sect. 4.2, we
  have not found these 8 sources deviating from the correlation
  between SFR and $\dot{M}$, which may imply X-ray emission from the
  jets is unimportant at least for these 8 sources.

    We find that our results are insensitive to any model
    parameters except $\delta$, which describes the fraction of the
    viscously dissipated energy directly heating the electrons in the accretion
    flow. In our calculations, $\delta=0.1$ is adopted, which is a
   typical value successfully used to model the observed spectra for some low-luminosity
   AGNs \citep*[e.g.,][]{qn99}. Although winds may be present in RIAFs, the detailed physics
   is still unclear and such winds are only described by an artificial
   power-law parameter $p_{\rm w}$ \citep*[e.g.,][]{bb99}. In RIAF
   spectral calculations for X-ray wavebands, the value $\delta$ is somewhat
   degenerate with $p_{\rm w}$ \citep*[e.g.,][]{qn99}. Our
   calculations show that similar conclusion can be obtained if
   $\delta=0.3$ and $p_{\rm w}=0.9$ are adopted for RIAF with winds.

      The present sample is a mixture of AGNs, of which both BH masses
    and bolometric luminosities are estimated. It is difficult to
    evaluate to what extent the sample is affected by the selection
    effects. A more robust sample is desired for testing this
    relation between $\dot{M}$ and SFR.

    \acknowledgments The authors are grateful to the referee for constructive suggestions on our
    paper. We would like to thank Jufu Lu, Feng Yuan,
  Weimin Gu, Shuangliang Li and Yiqing Lin for helpful
  discussions on the RIAF model and numerical calculations, and S.
  Satyapal, and R. P. Dudik,
  for useful discussions on their data. This work is supported by the
  National Science Fund for Distinguished Young Scholars (grant
  10325314) and the NSFC (grant 10333020).

\clearpage
\begin{deluxetable}{lcccrcc}
\tabletypesize{\scriptsize}
 \tablecaption{Re-estimated accretion rates of LINERs and Radio galaxies } \tablewidth{0pt} \tablehead{
\colhead{Galaxy Name} & \colhead{$\log_{10} \dot{M} ^{\rm\ a}$}&
\colhead{$\log_{10} \dot{M}^{\rm\ b}$} & \colhead{$\log_{10} SFR
^{\rm\ c}$} }

\startdata
$\rm LINERs ^{\rm\ d}$                             \\
\\
    NGC 4350    &     -5.35   &    -2.94  &   -0.83\\
    NGC 1052    &     -2.55   &    -1.39  &   -0.24\\
    NGC 3031    &     -4.05  &     -2.33  &   -0.76\\
    NGC 4278    &     -4.15  &     -1.73  &   -0.69\\
    NGC 4486    &     -3.65  &     -1.37  &   -0.83\\
    NGC 4579    &     -3.25  &     -2.04  &    0.09\\
    NGC 6500    &     -3.95  &     -1.89  &    0.01\\
    NGC 4203    &     -4.15  &     -2.31  &   -0.69\\
    NGC 4494    &     -5.35 &      -3.15  &   -1.22\\
    NGC 4594    &     -4.05 &      -1.80   &  -0.32\\
    NGC 4527    &     -5.45 &      -2.77   &   0.44\\
    NGC 4125    &     -5.55 &      -2.61   &  -0.62\\
    NGC 4374    &     -5.15  &     -1.99   &  -0.69\\
    NGC 4696    &     -4.05  &     -2.06   &  -0.47\\
    CGCG 162-010&     -2.35  &     -0.92   &   1.09\\
    IC  1459    &     -3.85  &     -1.74   &  -0.62\\
    NGC 2787    &     -5.95  &     -3.39   &  -1.16\\
    NGC  315    &     -3.00  &     -1.14  &    0.01\\
    NGC 2681    &     -5.30  &     -3.07  &   -0.47\\
    NGC 3169    &     -4.13   &    -2.32  &    0.35\\
    NGC 3245    &     -5.06  &     -2.56  &   -0.16\\
    NGC 3718    &     -3.78  &     -2.15  &   -0.54\\
    NGC 4258    &     -4.60  &     -2.94  &    0.01\\
    NGC 4261    &     -3.71  &     -1.73  &   -0.39\\
    NGC 4410A   &     -3.08  &     -1.38  &    0.27\\
    NGC 4457    &     -5.00   &    -3.21  &    0.01\\
    NGC 4552    &     -4.67   &    -2.33  &   -1.03\\
    NGC 4565    &     -5.13   &    -3.12  &   -0.16\\
    NGC 6482    &     -3.75   &    -1.75  &   -0.47\\
    3C  218     &     -2.40   &    -1.07  &    0.72\\

                                                      \\
$\rm 3C\ Radio\ galaxy ^{\rm\ d}$                             \\

\\
    3C  31    &       -3.751  &    -2.12  &    0.35\\
    3C  84    &       -1.751  &    -0.53   &   1.18\\
    3C  338   &       -3.451  &    -1.64   &   0.35\\
    3C  465    &      -3.151  &    -1.21  &    0.35\\
    3C  88     &      -3.351  &    -1.41  &    0.44\\
    3C  285    &      -4.251  &    -1.79  &    1.15\\
    3C  327    &      -3.351  &    -1.10  &    1.54\\
    3C  402    &      -3.351  &    -1.53  &   0.44\\

\enddata

\tablenotetext{a}{Result of \citet{sa05}, in unit of $\rm M_{\odot}
yr^{-1}$} \tablenotetext{b}{Our result, in unit of $\rm M_{\odot}
yr^{-1}$} \tablenotetext{c}{ in unit of $\rm M_{\odot} yr^{-1}$}
\tablenotetext{d}{The accretion rates of remaining 3 LINERs and 6
radio galaxies are estimated from $\dot{m}=L_{\rm bol}/L_{\rm Edd}$
with radiative efficiency $\epsilon_{r}^{0}=0.1$. Because the
$L_{X,\ 2-10\rm\ keV}$ of these sources are larger than the maximal
values predicted by RIAF model with parameters adopted in our paper
and RIAFs is assumed not present in these sources.}

\end{deluxetable}

\clearpage
\begin{figure}
\plotone {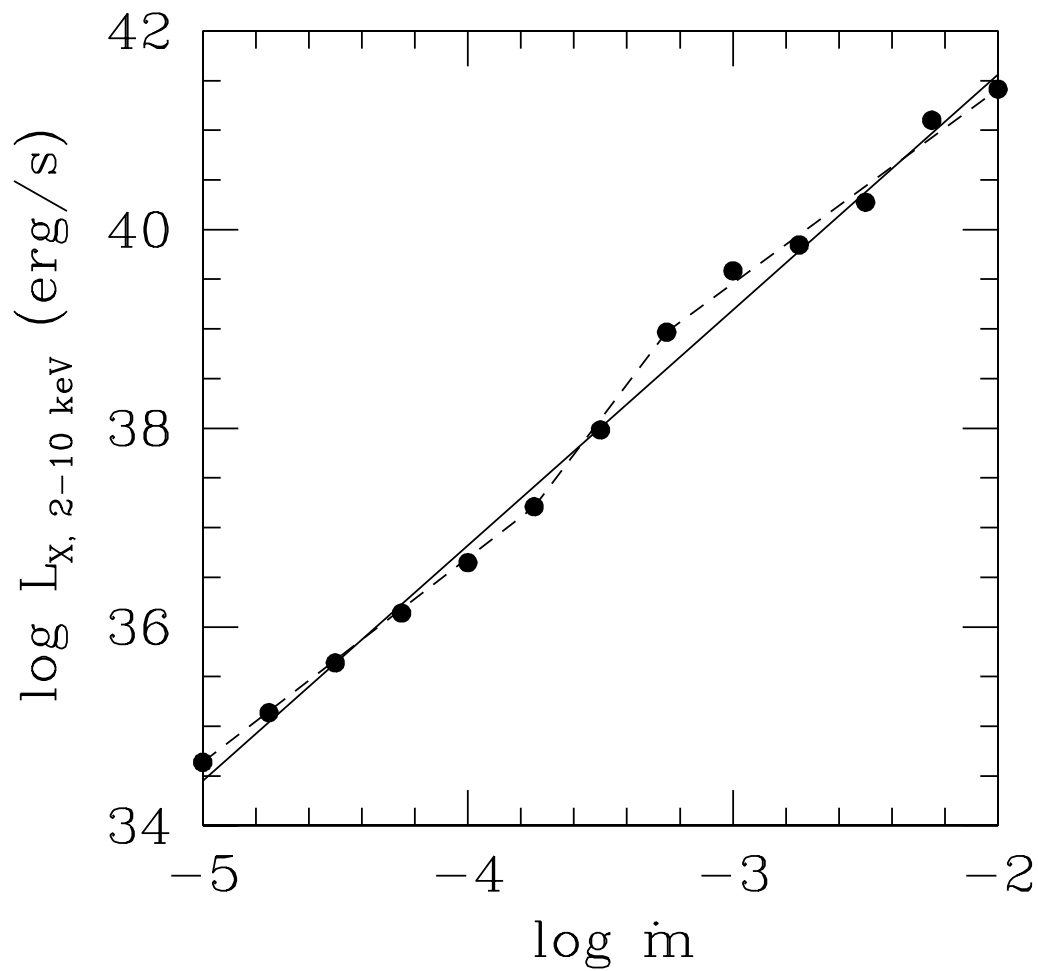}

 \caption{ The filled circles connected by the dashed line show the integrated 2-10 keV X-ray luminosity
 versus the dimensionless accretion rate $\dot{m}$ predicted by the RIAF model. The solid line
 is a linear fit of $L_{X,\ 2-10\ \rm keV}$ to the model calculated
points (filled circles). Model calculations are carried out for a
 $10^{8}M_{\odot}$ black hole, with $\alpha=0.2$, $\beta=0.8$ and
 $\delta=0.1$.
 \label{fig1}}

\end{figure}

\begin{figure}
\plotone {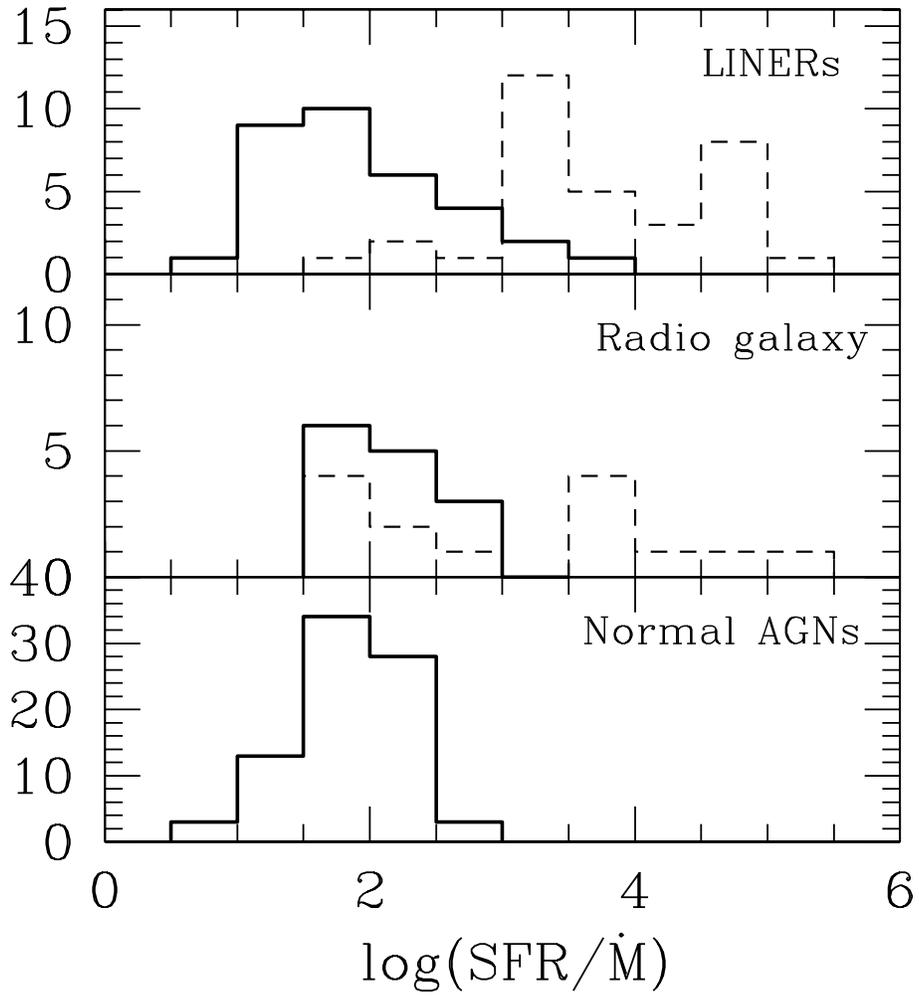}

 \caption{The histogram of SFR/$\dot{M}$ for the sources in the sample: 33 LINERs (the upper panel),
 14 radio galaxies (the middle panel), and 81 normal bright  AGNs (52 Seyferts, 15 quasars and
 14 NLS1s, the lower panel). The dashed lines in upper and middle
 panels are results of \citet{sa05}, while solid lines are ours.
 \label{fig2}}

\end{figure}

\begin{figure}
\plotone {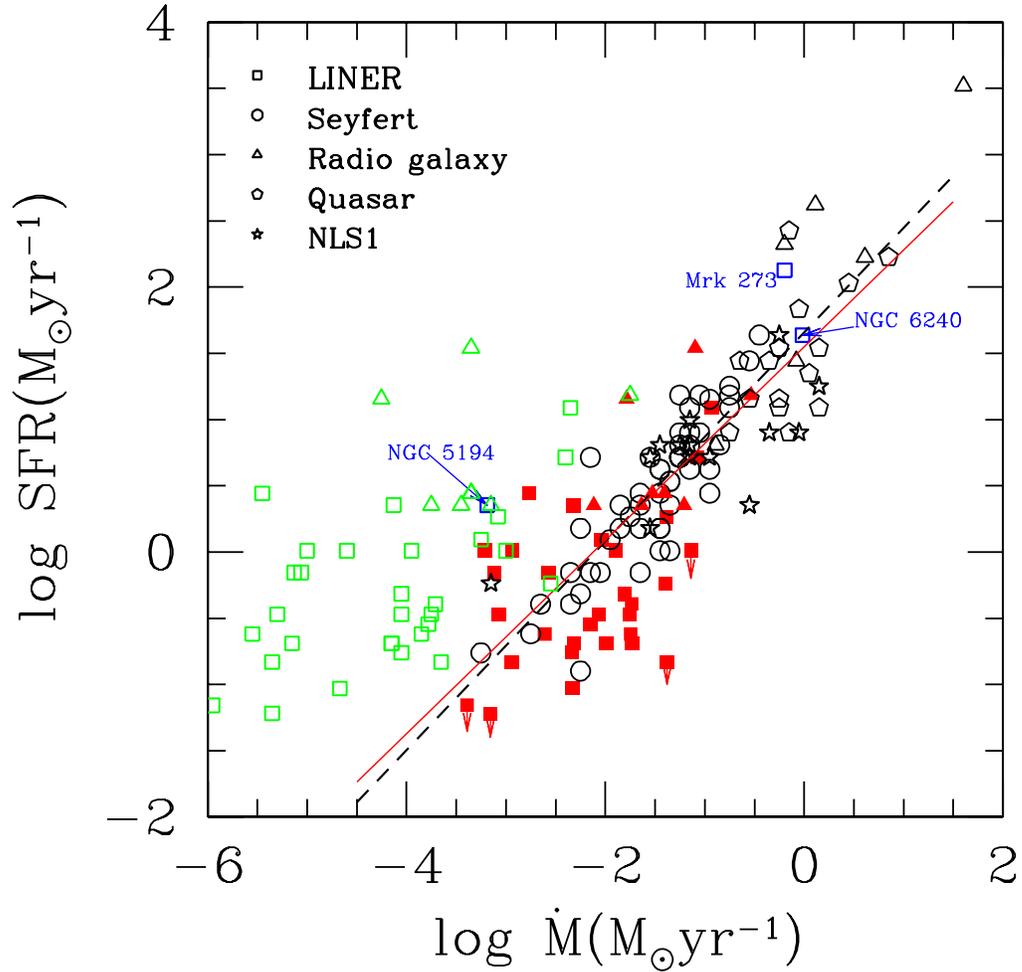}

 \caption{ Relation between SFR and $\dot{M}$ for the whole sample.
 The green-open squares are
 results of \citet{sa05}, while the red-solid squares are
 for the re-estimated accretion rates.
  The red-solid line represents the linear fit to the whole
  128 sources, while black-dashed line is for the normal bright AGNs.
 \label{fig3}}

\end{figure}

\end{document}